\documentclass[12pt,preprint]{aastex}
\shorttitle{Evolution in the Mean Type Ia Supernova Spectrum}
\shortauthors{Sullivan {\it et al.}}

\defcitealias{2008ApJ...674...51E}{E08}
\defcitealias{2007ApJ...659...98R}{R07}
\defcitealias{2008ApJ...684...68F}{F08}

\begin{document}

\newcommand\omatter{\ifmmode \Omega_{\mathrm{M}}\else $\Omega_{\mathrm{M}}$\fi}
\newcommand{\col}{\ensuremath{\mathcal{C}}}

\title{The Mean Type Ia Supernova Spectrum Over the Past 9 Gigayears}

\author {
  M.~Sullivan\altaffilmark{1},
  R.~S.~Ellis\altaffilmark{1,2}, 
  D.~A.~Howell\altaffilmark{3,4},
  A.~Riess\altaffilmark{5},
  P.~E.~Nugent\altaffilmark{6},
  A.~Gal-Yam\altaffilmark{7}
} 

\altaffiltext{1}{Department of Astrophysics, University of Oxford, Keble Road, Oxford OX1 3RH, UK}
\altaffiltext{2}{Department of Astrophysics, California Institute of Technology, MS 105-24, Pasadena, CA 91125 USA}
\altaffiltext{3}{Las Cumbres Observatory Global Telescope Network, 6740 Cortona Dr., Suite 102, Goleta, CA 93117}
\altaffiltext{4}{Department of Physics, University of California, Santa Barbara, Broida Hall, Mail Code 9530, Santa Barbara, CA 93106-9530}
\altaffiltext{5}{Department of Physics and Astronomy, Johns Hopkins University, 3400 N Charles St, Baltimore, MD 21218 USA}
\altaffiltext{6}{Lawrence Berkeley National Laboratory, 1 Cyclotron Rd., Berkeley CA 94720, USA}
\altaffiltext{7}{Astrophysics Group, Faculty of Physics, Weizmann Institute of Science, Rehovot 76100, Israel}

\email{sullivan@astro.ox.ac.uk}
 	
\begin{abstract}

  We examine the possibility of evolution with redshift in the mean
  rest-frame ultraviolet (UV; $\lambda\lesssim4500$\AA) spectrum of
  Type Ia Supernovae (SNe Ia) sampling the redshift range
  0$<$$z$$<$1.3.  We find new evidence for a decrease with redshift in
  the strength of intermediate-mass element (IME) features,
  particularly \ion{Si}{2} and to a lesser extent \ion{Ca}{2} `H\&K'
  and \ion{Mg}{2} blends, indicating lower IME abundances in the
  higher redshift SNe.  A larger fraction of luminous, wider
  light-curve width (higher ``stretch'') SNe Ia are expected at higher
  redshift than locally, so we compare our observed spectral evolution
  with that predicted by a redshift-evolving stretch distribution
  \citep{2007ApJ...667L..37H} coupled with a stretch-dependent SN Ia
  spectrum. We show that the sense of the spectral evolution can be
  reproduced by this simple model, though the highest redshift events
  seem additionally deficient in Si and Ca.  We also examine the mean
  SN Ia UV-optical colors as a function of redshift, thought to be
  sensitive to variations in progenitor composition. We find that the
  expected stretch variations are sufficient to explain the
  differences, although improved data at $z$$\sim$0 will enable more
  precise tests.  Thus, to the extent possible with the available
  datasets, our results support the continued use of SNe Ia as
  standardized candles.

\end{abstract}


\section{Introduction} 

Type Ia Supernovae (SNe Ia) have emerged as the most practical and
immediate path for improving our understanding of dark energy.
Techniques have been perfected for locating and studying hundreds of
events over the redshift range 0$<$$z$$<$1.5 where dark energy
manifests itself, and ambitious new facilities are being planned that
will take this progress to the next level of precision.  However, it
remains the case that the use of SNe Ia as cosmological probes is
purely empirical, exploiting relationships between photometric
properties such as peak luminosity, color and light curve width, that
result in distance estimates precise to $\simeq7\%$
\citep{2007A&A...466...11G,2007ApJ...659..122J,2008ApJ...681..482C}.

Early work by \citet{1995AJ....109....1H,1996AJ....112.2398H}
demonstrated that some of these photometric properties depend on their
host galaxy type, with SNe Ia in spirals possessing wider light curves
(higher ``stretch'') and are more luminous than those in older
elliptical hosts.  This has subsequently been categorized in terms of
(perhaps distinct) populations of `prompt' and `delayed' SNe Ia
\citep{2005A&A...433..807M,2006MNRAS.370..773M,2005ApJ...629L..85S},
whose properties appear to be governed by the star formation activity
of the host galaxy and therefore age of the progenitor stellar
population \citep{2001ApJ...554L.193H,2006ApJ...648..868S}.  With star
formation increasing at higher redshift, these demographic variations
produce a shift to more luminous, wider light curve events with
redshift \citep{2007ApJ...667L..37H}.  \citet{2008ApJ...684L..13S}
discuss the implications and possible biases introduced in dark energy
studies.

A less well-understood variation is the scatter in the rest-frame SN
Ia ultraviolet (UV) spectra from one event to another.  Although model
predictions differ in detail
\citep[e.g.][]{1998ApJ...495..617H,2000ApJ...530..966L,2008arXiv0803.0871S},
the form of the UV spectrum short-ward of 4000\AA\ is believed to be
sensitive to progenitor composition and explosion physics.  Given the
difficulty of studying this wavelength range in nearby events, the
most comprehensive studies have been accomplished using UV data
redshifted into the visible at $z\simeq$0.5
\citep{2008ApJ...674...51E,2008ApJ...684...68F}.  \citet[][hereafter
E08]{2008ApJ...674...51E} found significant scatter in the UV
continuum below $\simeq$4000\AA, at a level which exceeds that
predicted assuming quite significant metallicity variations in the
current models.

As both SN Ia progenitor age and metallicity likely evolve with
redshift, it is natural to ask whether any spectral variations
likewise appear redshift-dependent. Recent work by \citet[][hereafter
F08]{2008ApJ...684...68F} has tentatively identified a decrease with
redshift, at $\sim$2-$\sigma$ significance, in the strength of blend
of lines (\ion{Fe}{2}, \ion{Fe}{3}, \ion{Si}{2}) near 4800\AA\ in
maximum-light SN Ia spectra, which they attribute to \ion{Fe}{3}
5129\AA. In this letter, we examine the rest-frame
$\lambda$$\lesssim$4500\AA\ spectrum of SNe Ia over 0$<$$z$$<$1.3 by
combining data gathered using Keck at $z$$\simeq$0.5
\citepalias{2008ApJ...674...51E} and Hubble Space Telescope
(\textit{HST}) at $z$$>$1 \citep[][hereafter
R07]{2007ApJ...659...98R}.  Following \citet{2007ApJ...667L..37H}, our
goal is to investigate whether any spectral variations are present
that represent an evolutionary trend, and whether these can be
explained by the observed shift to more luminous events at higher
redshifts.

\section{SN Ia Spectral Samples}
\label{sec:samples}

We compile spectra from three sources corresponding to high,
intermediate and low redshift ($z$).  The intermediate-$z$ spectra are
described in \citetalias{2008ApJ...674...51E} together with all
aspects of their reduction.  Targets were drawn exclusively from the
Supernova Legacy Survey \citep[SNLS;][]{2006A&A...447...31A}, forming
a homogeneous sample representative of the SNLS parent sample. Host
galaxy contamination is removed using photometry of the hosts, with
the flux calibration corrected using contemporaneous imaging of the SN
light curves. This galaxy subtraction is a critical step when
examining ground-based spectra of high-redshift SNe, and an incorrect
continuum subtraction can systematically affect line feature strengths
and bias conclusions drawn from them; the wide wavelength range of
SNLS data allow an accurate host removal on individual spectra.  We
degrade the spectra to a resolution of 20\AA\ in the rest-frame to
match that of the high-$z$ \textit{HST} spectra discussed below (the
typical width of SN Ia spectral features is $\gtrsim100$\AA, or
$\gtrsim5$ resolution elements).

The $z$$>$0.9 spectra are taken from the \textit{HST} ACS grism
campaigns of \citet{2004ApJ...600L.163R} and
\citetalias{2007ApJ...659...98R}. Host galaxy subtraction is less
problematic as the SNe were resolved from their host due to the
superior resolution of \textit{HST}.  Any host that may lie underneath
the SN is subtracted by defining narrow ``sky'' regions above and
below the SN trace and interpolating to derive the contamination. The
resolution of the grism ($R=\lambda/\Delta\lambda\sim200$ at 8000\AA)
equates to $\sim20$\AA\ in the rest-frame at 4000\AA\ for a $z$=1 SN,
improving for higher-$z$ events. We rebin all spectra to the 20\AA\
wavelength scale.

At low-$z$, samples are heterogeneous in nature. Both the
\citetalias{2007ApJ...659...98R} and \citetalias{2008ApJ...674...51E}
samples were found with ``rolling searches'' and can be considered
complete with regard to host galaxy and environment; the same is not
true for local searches which frequently involve targeting known
galaxies. We replace the sample used in
\citetalias{2008ApJ...674...51E} with that of \citet[][hereafter
M08]{2008AJ....135.1598M}, which has the benefit of a homogeneous
reduction technique, resulting in spectra that are close to
spectrophotomteric (M08). For each SN we select the closest spectrum
to a phase of +3d (the mean of the higher-$z$ samples). Due to the
large apparent size of the host galaxies, the surface brightness at
the SN position is small.  The M08 spectra are optical only, so at
$\lambda$$\lesssim$3600\AA, we supplement the sample with the
space-based UV spectra of SN1992A \citep{1993ApJ...415..589K}, SN1981B
\citep{1983ApJ...270..123B} and SN2001ba
\citep[e.g.][]{2008ApJ...686..117F}, the only three spectroscopically
normal SNe Ia in the appropriate phase range.

\begin{figure}
\plotone{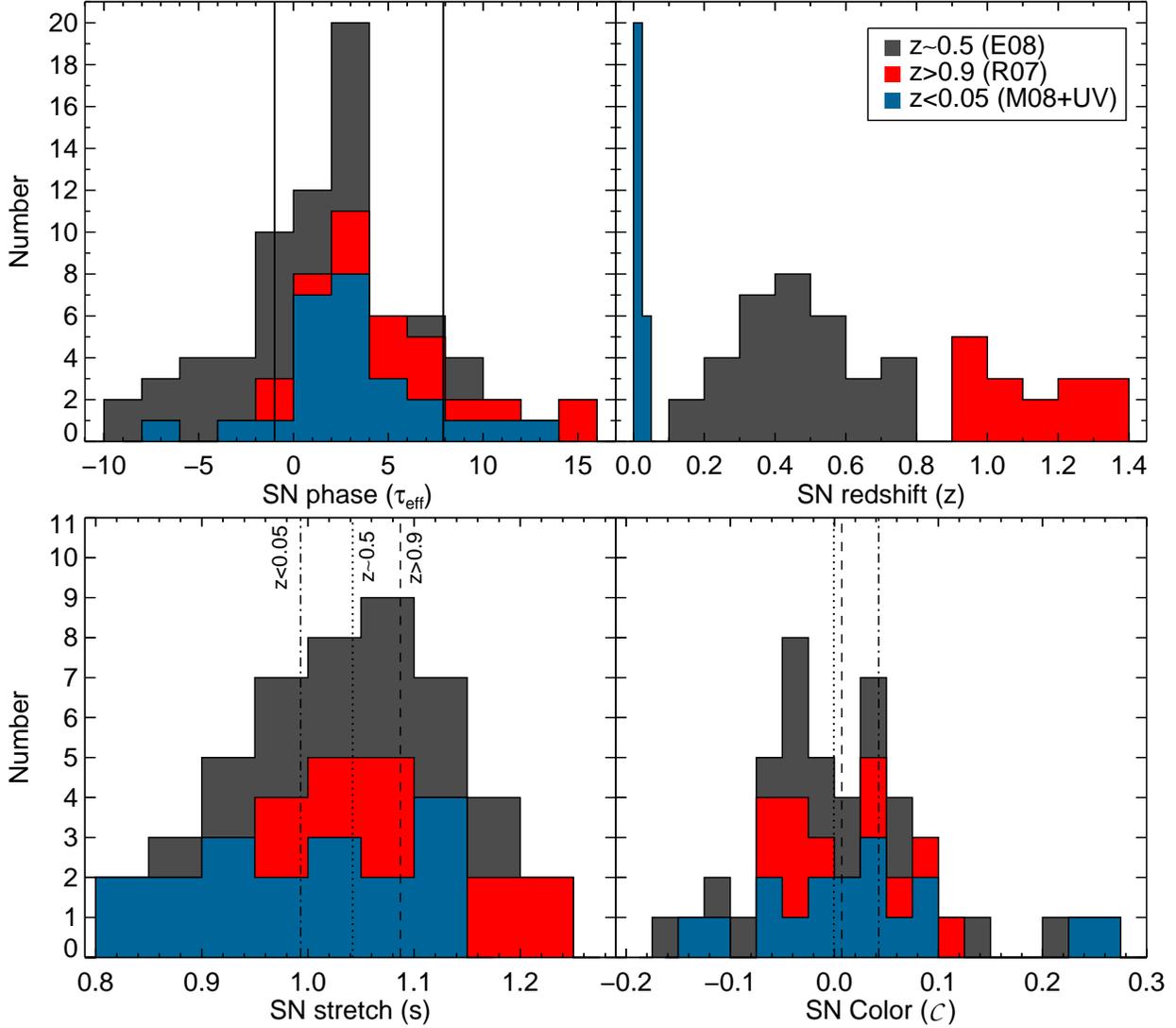}
\caption{The distribution of various properties of the spectral
  sample. Upper left: The phase distribution of the entire sample. The
  vertical lines show the limits used to define the comparison
  samples. Upper right: The redshift distribution of the entire
  sample. Lower left: The SN stretch distribution for just the
  comparison sample. The dotted/dashed/dot-dash vertical lines show
  the weighted means for the SNLS/low-$z$/HST samples.  Lower right:
  The SN color distributions, weighted mean colors
  marked.\label{fig:tzs_hist}}
\end{figure}

The light curves of all the SNe Ia are fit using the SiFTO light curve
fitter \citep{2008ApJ...681..482C} to place their photometric
parameters on the same system. The distributions in light curve phase
($\tau_{\mathrm{eff}}$; measured in days relative to maximum light in
the rest-frame $B$ band divided by light curve stretch), $z$, stretch
($s$) and rest-frame $B-V$ color (\col), are shown in
Figure~\ref{fig:tzs_hist}.  The high-$z$ \textit{HST} spectra are
typically at a later phase due to the scheduling and search strategy
used. Note that the SNLS, a rolling search, produces more early SN Ia
spectra than more traditional search methods used at low-$z$. We
restrict our analysis to a comparison sample of spectra with
-1$<$$\tau_{\mathrm{eff}}$$<$8 (important given the evolution in SN
spectra with light curve phase), a stretch range of 0.8$<$$s$$<$1.30,
and a color range of -0.2$<$$\col$$<$0.35. We only consider
``spectroscopically normal'' SNe, excluding obvious, low-Si
SN1991T-like events \citep{1992ApJ...384L..15F}. The mean properties
are given in Table~\ref{tab:sampleprops}.

\begin{deluxetable}{ccccc}
  \tablecaption{Mean properties of the SN Ia samples}
  \tablehead{\colhead{$\overline{z}$} & \colhead{N} &
    \colhead{$\overline{\tau}_{\mathrm{eff}}$} &
    \colhead{$\overline{s}$} & \colhead{$\overline{\col}$}} \startdata
  1.16 (0.97 to 1.39)  & 12 & 3.5 & 1.087(23) & 0.007(18)\\
  0.48 (0.12 to 0.73)  & 18 & 2.0 & 1.042(20) & 0.000(20)\\
  0.02 (0.00 to 0.04)  & 18 & 2.8 & 0.993(25) & 0.042(28)\\
0.01 (UV spectra only) &  3 & 4.1 & 0.91(06)  & -0.01(0.05) \\
\enddata
\label{tab:sampleprops}
\end{deluxetable}

The stretch distributions are marginally different, with evidence for
an increase in mean stretch with redshift at $\sim1.5\sigma$
significance, as expected following the analysis of
\citet{2007ApJ...667L..37H}. The \col\ distributions of all three
samples are consistent, arguing \textit{against} the stretch
differences arising primarily from selection effects related to search
biases, as these would also be manifest in the color histograms. The
sample of low-$z$ SNe with UV spectra has a particularly low mean
stretch, with two of these SNe at the lower end of the stretch range
(SN1992a at 0.82 and SN1981b at 0.89).

\section{Spectral Comparisons}
\label{sec:spectral-comparisons}

We construct mean spectra for the three samples following
\citetalias{2008ApJ...674...51E}. All spectra are corrected for Milky
Way extinction and color-corrected using the SALT-2 color law
\citep{2007A&A...466...11G}, reducing, though not eliminating,
dispersion among SN Ia spectra and aligning them in the mean
\citepalias{2008ApJ...674...51E}.  The spectra are normalized to have
the same flux through a box filter defined from rest-frame 3750 to
4100\AA\ (denoted $U$'), chosen as a common wavelength region for all
spectra (changing this wavelength range does not change our results).
The error on the mean spectrum is estimated using a bootstrap
resampling technique \citepalias[see][for more
details]{2008ApJ...674...51E}. \citetalias{2007ApJ...659...98R}
previously presented a mean spectrum for the high-$z$ \textit{HST}
spectra (used in comparisons with lower-redshift by
\citetalias{2008ApJ...684...68F}), calculated over a wider phase range
with different error estimation techniques, and lacking
color-corrections on the individual spectra. A comparison of this
\citetalias{2007ApJ...659...98R} mean with (e.g.)  the
intermediate-$z$ mean of \citetalias{2008ApJ...674...51E} would not be
appropriate, so we rederive the \textit{HST} mean for comparison on a
consistent basis\footnote{The \citetalias{2007ApJ...659...98R} SNe
  used are: HST05Str, HST05Gab, HST05Lan, HST04Pat, HST04Gre,
  HST04Eag, HST04Sas, SN2003az(Tor), SN2002fw(Aph), SN2003dy(Bor),
  HST04Omb, sn2002dd.}.

\begin{figure}
\plotone{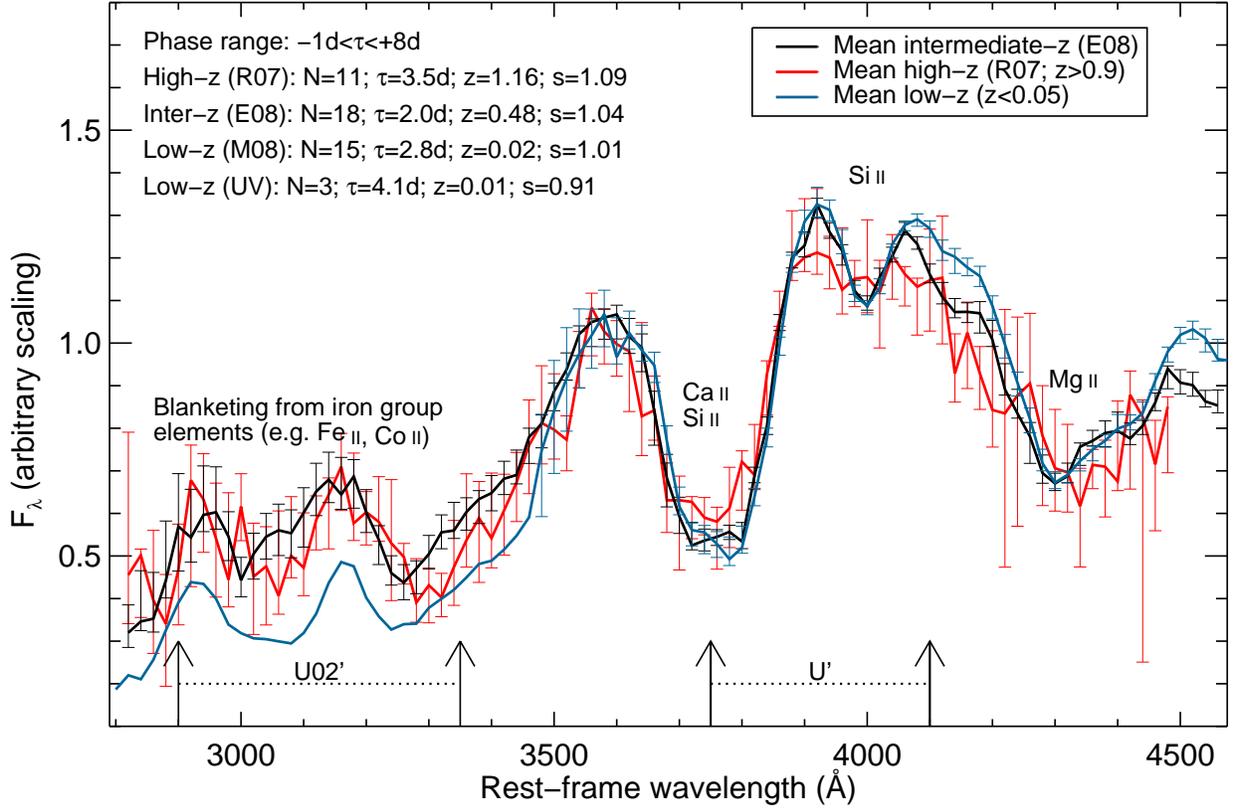}
\caption{The comparison of the Keck-SNLS (black), low-$z$ (blue) and
  \textit{HST} (red) mean SN~Ia spectra. The resolution is 20\AA\ and
  no additional smoothing is performed. The error bars are derived
  using a bootstrap resampling technique \citep[see][for
  details]{2008ApJ...674...51E}. At $\lambda\lesssim3500$\AA\ in the
  low-$z$ SNe, no error bars are plotted as only three SNe enter the
  sample. The vertical arrows show the boundaries of the $U$' (right)
  and $U02$' (left) filters.
  \label{fig:mean_spectrum}}
\end{figure}

The comparison of the mean spectra is shown in
Figure~\ref{fig:mean_spectrum}. As noted by previous authors
conducting similar analyses, the spectra appear very similar. In
particular, with the exception of the low-$z$ UV mean spectrum, the
agreement in the shape of the continua is remarkable, suggesting the
color corrections made to the spectra based on their optical color are
effective in aligning the spectra across the entire wavelength and $z$
range. However, we find new evidence for variations in some ion
feature strengths, most obviously around 4000\AA\ at the position of
the blueshifted \ion{Si}{2} 4128\AA\ feature.  Since the luminosity of
the SN Ia is controlled by the amount of $^{56}$Ni synthesized in the
explosion, for a Chandrasekhar-mass explosion, brighter explosions
imply increased $^{56}$Ni production at the expense of intermediate
mass elements (IMEs)\footnote{Stable, nuclear statistical equilibrium
  elements like $^{54}$Fe or $^{58}$Ni play a role, but they are
  relatively constant from SN to SN
  \citep[e.g.][]{2007Sci...315..825M}, and have only a small effect on
  this argument.} such as Si, Ca and Mg.  So to first order one would
expect to see fewer IMEs in more luminous, or broader-lightcurve SNe.
Indeed, the \ion{Si}{2} 4128 feature has been shown to have a lower
equivalent width in broader-lightcurve SNe Ia
\citep{2007A&A...466...11G,2008ApJ...674...51E} and, by extension, in
SNe Ia originating in spiral galaxies \citep{2008A&A...477..717B}.
Given the observed drift in mean stretch with $z$ of the SN Ia
population \citep{2007ApJ...667L..37H}, arising from shifting
demographics of the SN Ia population, a generic prediction of
two-component prompt+delayed SN Ia rate models is that the amount of
IMEs in general, and Si in particular, in the average high-$z$
spectrum should be smaller \citep{2007ApJ...667L..37H}. This is in the
same sense as the trends observed in the comparison of the means in
Figure~\ref{fig:mean_spectrum}.

The interpretation of SN Ia spectra is complicated, and this simple
picture may not tell the whole story.  Observing a spectral feature in
absorption only constrains the amount of material in a given
ionization state, along the line of sight, above the photosphere at
the time of the spectrum.  At maximum light, approximately 2/3 of the
mass of the SN Ia remains hidden under the photosphere
\citep[e.g.][]{1998ApJ...495..617H}.  One can imagine a pathological
situation in which more luminous SNe Ia have fewer IMEs above the
photosphere, and more below, but this would be at odds with the basic
physical principle that fusion in denser regions produces heavier
elements, a generic property of SN Ia models.  Instead, a more likely
complication is that more luminous SNe are hotter, and thus more
ionized.  In this situation, in the hottest regions \ion{Si}{2} may be
ionized to \ion{Si}{3}, resulting in lower \ion{Si}{2} equivalent
widths (EWs).  Fortunately, this potential complication does not
change the basic principle that lower EWs of singly ionized IMEs are
expected in the spectra of more luminous SNe Ia.

\citetalias{2008ApJ...684...68F}, comparing $z$=0 and $z$=0.2--0.8
mean spectra, find a lower strength in the \ion{Fe}{2} blend at
4800\AA\ in higher-$z$ spectra. \citetalias{2008ApJ...684...68F} state
they are unable to distinguish evolving SN Ia demographics with
evolution in the SN properties. At first sight, their trend appears
consistent with the results of this paper: higher-stretch SNe Ia will
be hotter, and thus Fe ionized to \ion{Fe}{3} at the expense of
\ion{Fe}{2}. However, the 4800\AA\ feature is a complex one, a blend
of \ion{Si}{2} (5041\AA, 5055\AA), \ion{Fe}{2} (4923\AA, 5018\AA,
5169\AA) and \ion{Fe}{3} (5129\AA)
\citep[e.g.][]{1999ApJS..121..233H}.  In fact,
\citetalias{2008ApJ...684...68F} attribute the weakening strength of
the red side of the feature to \ion{Fe}{3} which would be inconsistent
with the simple picture outlined above. Given the presence of the
other \ion{Fe}{2} and \ion{Si}{2} lines in this 4800\AA\ feature,
further detailed modeling work is required before a robust conclusion
with regards to this feature can be reached.

\begin{figure}
\plotone{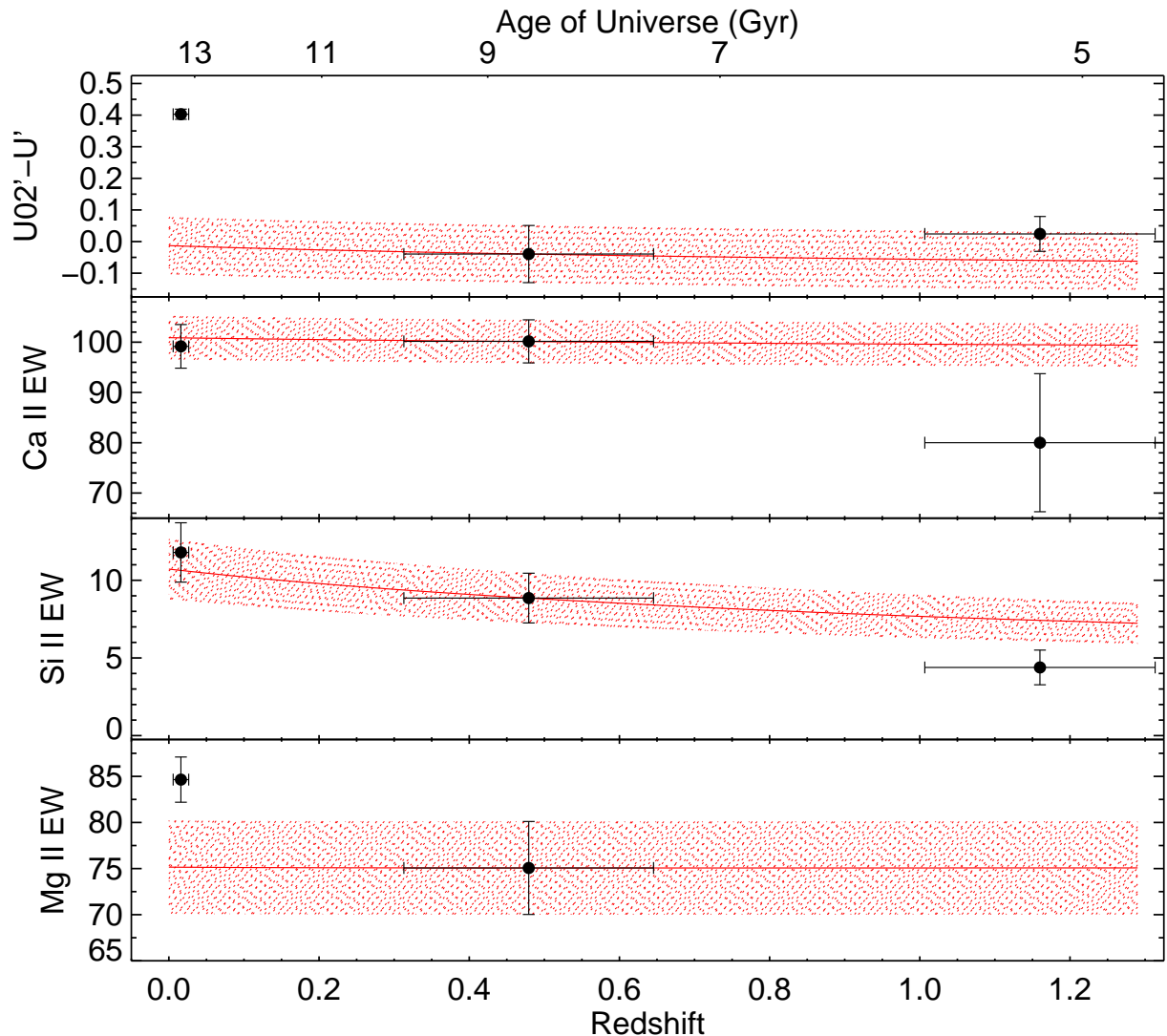}
\caption{The variation with redshift of various features in the three
  mean spectra. Top panel shows the $U02$'-$U$' color measured from
  the spectra, and the lower three panels show the equivalent width
  (EW) of the three features \ion{Ca}{2}, \ion{Si}{2} and \ion{Mg}{2}.
  Redshift errors are the standard deviations of each sample.
  Overplotted in red is the predicted variation from the observed
  shift in SN demographics (stretch) with redshift, normalized to the
  intermediate-$z$ sample; the hashed area shows the uncertainty in
  this prediction based on the errors in the intermediate-$z$
  measurement.\label{fig:features_z}}
\end{figure}

We investigate the variations in more detail by measuring the strength
of three accessible IME spectral features (the blend of \ion{Ca}{2} at
rest-frame $\sim$3950\AA\ and \ion{Si}{2} at $\sim$3860\AA,
\ion{Si}{2} at 4128\AA, and \ion{Mg}{2} at 4481\AA). We fit a Gaussian
on a linear pseudo-continuum background and calculate the (pseudo) EW
\citep[e.g.][]{2007A&A...470..411G}. The error is estimated by
performing the EW fit on each of the bootstrap resampled spectra and
taking the standard deviation of the resulting distribution. The
variation with redshift of these features is shown in
Figure~\ref{fig:features_z}; the trends are for a decreasing EW for
all three ions with increasing redshift. We emphasize that these
measurements are made on mean spectra; in many cases the individual
high-$z$ spectra show evidence of \ion{Si}{2}
\citepalias{2007ApJ...659...98R}.

Further into the UV, the continuum level shows a redshift-dependent
trend, with the spectra broadly consistent from 0.5$<$$z$$<$1, but
depressed at $z$$<$0.05. Potentially this is an important result;
however the different SN Ia samples possess different mean stretches,
and the UV-optical color of the higher-stretch samples may be expected
to appear bluer
\citep[e.g.][]{2007A&A...466...11G,2008ApJ...674...51E}. Furthermore
the paucity of local data and the intrinsic UV scatter from one event
to another \citepalias{2008ApJ...674...51E} may explain such
discrepancies. We quantify this ``evolution'' by measuring a
UV-optical color on the mean spectra with two synthetic filters
denoted $U02$' \citep[following][]{2008ApJ...681..482C} and $U$', both
marked in Figure~\ref{fig:mean_spectrum}.  Errors are again estimated
by analyzing the bootstrapped spectra, and the results are plotted in
Figure~\ref{fig:features_z}.

\section{Discussion}
\label{sec:discussion}

A key question is whether the changes seen in
Figures~\ref{fig:mean_spectrum} and \ref{fig:features_z} are a
signature of evolution in individual SN Ia properties, or instead a
product of changing demographics in the SN population, i.e. the
relative increase in SNe Ia from younger progenitors at higher
redshift. Are the spectral properties of SNe Ia, at a fixed stretch,
evolving with redshift, or is the previously observed redshift drift
in the mean stretch causing the observed spectral evolution? To
address this, we predict how the various spectral features might be
expected to evolve with redshift were a demographic change solely
responsible. We use the analysis of \citet{2007ApJ...667L..37H}, which
predicts how the fraction of the two SN Ia components (e.g., `A+B',
delayed+prompt, or old+young), and hence the mean SN Ia stretch,
changes with redshift. This work is based on the galaxy-type dependent
stretch distributions of \citet{2006ApJ...648..868S} coupled with a
cosmic star formation history and various simple two-component SN Ia
rate models \citep{2005ApJ...629L..85S,2006MNRAS.370..773M}.

We construct a mean spectrum for low and high-stretch SNe Ia based on
the sample of \citetalias{2008ApJ...674...51E}, defining low-stretch
($s\leq s_0$) and high-stretch ($s>s_0$) sub groups. We use a split
point of $s_0=1.0$ \citep[e.g.][]{2007ApJ...667L..37H}, giving
$\overline{s}_{\mathrm{low}}=0.92$ and
$\overline{s}_{\mathrm{high}}=1.09$. The low-stretch mean is assigned
to the ``A'' (delayed) component, and the high-stretch mean to the
``B'' (prompt) component. A model mean spectrum for any redshift can
then be constructed by combining the two stretch-dependent means in
the appropriate ratio of the two components.  Predictions for the
behavior of the various spectral features can be made by measuring
these features on mean spectra constructed at different redshifts, and
are over-plotted on Figure~\ref{fig:features_z} (there is some
circularity here, as the sample of spectra are also used to make the
intermediate-$z$ measurement; however here we are interested in the
predicted \textit{change} with redshift rather than the absolute
values). The sensitivity of the predictions to the stretch split point
is explored varying $s_0$ by 5\%; this has a very minor effect on the
results.

The predicted tracks of the model generally match the observations,
although the uncertainties remain large. The decrease in \ion{Si}{2}
EW with redshift is reproduced, though there is evidence that the
high-$z$ mean shows an additional deficit in both Si and Ca compared
to the lower-redshift spectra. This could also reflect an increase in
the fraction of SN1991T-like events, believed to arise from a younger
population, in the $z$$>$1 sample, where excluding these objects is
more difficult due to the lower signal-to-noise of the spectra. The
predicted change in UV-optical color fails to match the data, though
the robust datasets
\citepalias{2008ApJ...674...51E,2007ApJ...659...98R} show only modest
changes consistent with that implied by the change in mean stretch.
The main discrepancy occurs with the local data. Such a discontinuous
behavior seems unlikely and probably arises from the incomplete sample
of local UV spectra, re-emphasizing the need for local,
spectroscopically-normal, SNe Ia observed in the UV.

We have shown, for the first time, that the strength of some SN Ia
intermediate-mass element spectral features evolve systematically with
redshift. However, we hypothesize that this evolution is related to
the previously identified shift in SN Ia photometric properties with
redshift \citep{2007ApJ...667L..37H}, and we show it can be partially
modeled using a simple two-component SN Ia rate model and the stretch
dependent spectra of \citetalias{2008ApJ...674...51E}.  After
accounting for these expected trends, some small discrepancies in the
highest redshift spectra remain, though none at high significance.
Thus, we conclude that the observed evolution in SN Ia spectral
features can be explained by, and is consistent with, changes in SN Ia
demographics, rather than evolution in the properties of individual,
photometrically-similar SNe. Controlling for these demographic shifts
is the challenge for SN Ia precision cosmology experiments, for
example by ensuring that an appropriate $k$-correction template is
used as a function of redshift, and confirming that the empirical
relationships used to calibrate SNe Ia in cosmological applications
are valid across all environments.

\acknowledgements 

MS and RSE acknowledge support from the Royal Society.  PEN
acknowledges support from the US Department of Energy Scientific
Discovery through Advanced Computing program under contract
DE-FG02-06ER06-04. AG acknowledges support by the Benoziyo Center for
Astrophysics, a research grant from Peter and Patricia Gruber Awards,
and the William Z. and Eda Bess Novick New Scientists Fund at the
Weizmann Institute.


\begin{thebibliography}{32}
\expandafter\ifx\csname natexlab\endcsname\relax\def\natexlab#1{#1}\fi

\bibitem[{{Astier} {et~al.}(2006){Astier}, {Guy}, {Regnault}, {Pain},
  {Aubourg}, {Balam}, {Basa}, {Carlberg}, {Fabbro}, {Fouchez}, {Hook},
  {Howell}, {Lafoux}, {Neill}, {Palanque-Delabrouille}, {Perrett}, {Pritchet},
  {Rich}, {Sullivan}, {Taillet}, {Aldering}, {Antilogus}, {Arsenijevic},
  {Balland}, {Baumont}, {Bronder}, {Courtois}, {Ellis}, {Filiol}, {Gon{\c
  c}alves}, {Goobar}, {Guide}, {Hardin}, {Lusset}, {Lidman}, {McMahon},
  {Mouchet}, {Mourao}, {Perlmutter}, {Ripoche}, {Tao}, \&
  {Walton}}]{2006A&A...447...31A}
{Astier}, P., et al. 2006, \aap, 447, 31

\bibitem[{{Branch} {et~al.}(1983){Branch}, {Lacy}, {McCall}, {Sutherland},
  {Uomoto}, {Wheeler}, \& {Wills}}]{1983ApJ...270..123B}
{Branch}, D., {Lacy}, C.~H., {McCall}, M.~L., {Sutherland}, P.~G., {Uomoto},
  A., {Wheeler}, J.~C., \& {Wills}, B.~J. 1983, \apj, 270, 123

\bibitem[{{Bronder} {et~al.}(2008){Bronder}, {Hook}, {Astier}, {Balam},
  {Balland}, {Basa}, {Carlberg}, {Conley}, {Fouchez}, {Guy}, {Howell}, {Neill},
  {Pain}, {Perrett}, {Pritchet}, {Regnault}, {Sullivan}, {Baumont}, {Fabbro},
  {Filliol}, {Perlmutter}, \& {Ripoche}}]{2008A&A...477..717B}
{Bronder}, T.~J., et al. 2008, \aap, 477, 717

\bibitem[{{Conley} {et~al.}(2008){Conley}, {Sullivan}, {Hsiao}, {Guy},
  {Astier}, {Balam}, {Balland}, {Basa}, {Carlberg}, {Fouchez}, {Hardin},
  {Howell}, {Hook}, {Pain}, {Perrett}, {Pritchet}, \&
  {Regnault}}]{2008ApJ...681..482C}
{Conley}, A., et al. 2008, \apj, 681, 482

\bibitem[{{Ellis} {et~al.}(2008){Ellis}, {Sullivan}, {Nugent}, {Howell},
  {Gal-Yam}, {Astier}, {Balam}, {Balland}, {Basa}, {Carlberg}, {Conley},
  {Fouchez}, {Guy}, {Hardin}, {Hook}, {Pain}, {Perrett}, {Pritchet}, \&
  {Regnault}}]{2008ApJ...674...51E}
{Ellis}, R.~S., et al. 2008, \apj, 674, 51

\bibitem[{{Filippenko} {et~al.}(1992){Filippenko}, {Richmond}, {Matheson},
  {Shields}, {Burbidge}, {Cohen}, {Dickinson}, {Malkan}, {Nelson}, {Pietz},
  {Schlegel}, {Schmeer}, {Spinrad}, {Steidel}, {Tran}, \&
  {Wren}}]{1992ApJ...384L..15F}
{Filippenko}, A.~V., et al. 1992, \apjl, 384, L15

\bibitem[{{Foley} {et~al.}(2008{\natexlab{a}}){Foley}, {Filippenko},
  {Aguilera}, {Becker}, {Blondin}, {Challis}, {Clocchiatti}, {Covarrubias},
  {Davis}, {Garnavich}, {Jha}, {Kirshner}, {Krisciunas}, {Leibundgut}, {Li},
  {Matheson}, {Miceli}, {Miknaitis}, {Pignata}, {Rest}, {Riess}, {Schmidt},
  {Smith}, {Sollerman}, {Spyromilio}, {Stubbs}, {Suntzeff}, {Tonry},
  {Wood-Vasey}, \& {Zenteno}}]{2008ApJ...684...68F}
{Foley}, R.~J., et al. 2008{\natexlab{a}},
  \apj, 684, 68

\bibitem[{{Foley} {et~al.}(2008{\natexlab{b}}){Foley}, {Filippenko}, \&
  {Jha}}]{2008ApJ...686..117F}
{Foley}, R.~J., {Filippenko}, A.~V., \& {Jha}, S.~W. 2008{\natexlab{b}}, \apj,
  686, 117

\bibitem[{{Garavini} {et~al.}(2007){Garavini}, {Folatelli}, {Nobili},
  {Aldering}, {Amanullah}, {Antilogus}, {Astier}, {Blanc}, {Bronder}, {Burns},
  {Conley}, {Deustua}, {Doi}, {Fabbro}, {Fadeyev}, {Gibbons}, {Goldhaber},
  {Goobar}, {Groom}, {Hook}, {Howell}, {Kashikawa}, {Kim}, {Kowalski},
  {Kuznetsova}, {Lee}, {Lidman}, {Mendez}, {Morokuma}, {Motohara}, {Nugent},
  {Pain}, {Perlmutter}, {Quimby}, {Raux}, {Regnault}, {Ruiz-Lapuente},
  {Sainton}, {Schahmaneche}, {Smith}, {Spadafora}, {Stanishev}, {Thomas},
  {Walton}, {Wang}, {Wood-Vasey}, \& {Yasuda}}]{2007A&A...470..411G}
{Garavini}, G., et al. 2007, \aap, 470, 411

\bibitem[{{Guy} {et~al.}(2007){Guy}, {Astier}, {Baumont}, {Hardin}, {Pain},
  {Regnault}, {Basa}, {Carlberg}, {Conley}, {Fabbro}, {Fouchez}, {Hook},
  {Howell}, {Perrett}, {Pritchet}, {Rich}, {Sullivan}, {Antilogus}, {Aubourg},
  {Bazin}, {Bronder}, {Filiol}, {Palanque-Delabrouille}, {Ripoche}, \&
  {Ruhlmann-Kleider}}]{2007A&A...466...11G}
{Guy}, J., et al. 2007, \aap, 466, 11

\bibitem[{{H{\" o}flich} {et~al.}(1998){H{\" o}flich}, {Wheeler}, \&
  {Thielemann}}]{1998ApJ...495..617H}
{H{\" o}flich}, P., {Wheeler}, J.~C., \& {Thielemann}, F.~K. 1998, \apj, 495,
  617

\bibitem[{{Hamuy} {et~al.}(1995){Hamuy}, {Phillips}, {Maza}, {Suntzeff},
  {Schommer}, \& {Aviles}}]{1995AJ....109....1H}
{Hamuy}, M., {Phillips}, M.~M., {Maza}, J., {Suntzeff}, N.~B., {Schommer},
  R.~A., \& {Aviles}, R. 1995, \aj, 109, 1

\bibitem[{{Hamuy} {et~al.}(1996){Hamuy}, {Phillips}, {Suntzeff}, {Schommer},
  {Maza}, \& {Aviles}}]{1996AJ....112.2398H}
{Hamuy}, M., {Phillips}, M.~M., {Suntzeff}, N.~B., {Schommer}, R.~A., {Maza},
  J., \& {Aviles}, R. 1996, \aj, 112, 2398

\bibitem[{{Hatano} {et~al.}(1999){Hatano}, {Branch}, {Fisher}, {Millard}, \&
  {Baron}}]{1999ApJS..121..233H}
{Hatano}, K., {Branch}, D., {Fisher}, A., {Millard}, J., \& {Baron}, E. 1999,
  \apjs, 121, 233

\bibitem[{{Howell}(2001)}]{2001ApJ...554L.193H}
{Howell}, D.~A. 2001, \apjl, 554, L193

\bibitem[{{Howell} {et~al.}(2007){Howell}, {Sullivan}, {Conley}, \&
  {Carlberg}}]{2007ApJ...667L..37H}
{Howell}, D.~A., {Sullivan}, M., {Conley}, A., \& {Carlberg}, R. 2007, \apjl,
  667, L37

\bibitem[{{Jha} {et~al.}(2007){Jha}, {Riess}, \&
  {Kirshner}}]{2007ApJ...659..122J}
{Jha}, S., {Riess}, A.~G., \& {Kirshner}, R.~P. 2007, \apj, 659, 122

\bibitem[{{Kirshner} {et~al.}(1993){Kirshner}, {Jeffery}, {Leibundgut},
  {Challis}, {Sonneborn}, {Phillips}, {Suntzeff}, {Smith}, {Winkler}, {Winge},
  {Hamuy}, {Hunter}, {Roth}, {Blades}, {Branch}, {Chevalier}, {Fransson},
  {Panagia}, {Wagoner}, {Wheeler}, \& {Harkness}}]{1993ApJ...415..589K}
{Kirshner}, R.~P., et al. 1993, \apj, 415,
  589

\bibitem[{{Lentz} {et~al.}(2000){Lentz}, {Baron}, {Branch}, {Hauschildt}, \&
  {Nugent}}]{2000ApJ...530..966L}
{Lentz}, E.~J., {Baron}, E., {Branch}, D., {Hauschildt}, P.~H., \& {Nugent},
  P.~E. 2000, \apj, 530, 966

\bibitem[{{Mannucci} {et~al.}(2006){Mannucci}, {Della Valle}, \&
  {Panagia}}]{2006MNRAS.370..773M}
{Mannucci}, F., {Della Valle}, M., \& {Panagia}, N. 2006, \mnras, 370, 773

\bibitem[{{Mannucci} {et~al.}(2005){Mannucci}, {della Valle}, {Panagia},
  {Cappellaro}, {Cresci}, {Maiolino}, {Petrosian}, \&
  {Turatto}}]{2005A&A...433..807M}
{Mannucci}, F., et al. 2005, \aap, 433, 807

\bibitem[{{Matheson} {et~al.}(2008){Matheson}, {Kirshner}, {Challis}, {Jha},
  {Garnavich}, {Berlind}, {Calkins}, {Blondin}, {Balog}, {Bragg}, {Caldwell},
  {Dendy Concannon}, {Falco}, {Graves}, {Huchra}, {Kuraszkiewicz}, {Mader},
  {Mahdavi}, {Phelps}, {Rines}, {Song}, \& {Wilkes}}]{2008AJ....135.1598M}
{Matheson}, T., et al. 2008, \aj, 135,
  1598

\bibitem[{{Mazzali} {et~al.}(2007){Mazzali}, {R{\"o}pke}, {Benetti}, \&
  {Hillebrandt}}]{2007Sci...315..825M}
{Mazzali}, P.~A., {R{\"o}pke}, F.~K., {Benetti}, S., \& {Hillebrandt}, W. 2007,
  Science, 315, 825

\bibitem[{{Riess} {et~al.}(2004){Riess}, {Strolger}, {Tonry}, {Tsvetanov},
  {Casertano}, {Ferguson}, {Mobasher}, {Challis}, {Panagia}, {Filippenko},
  {Li}, {Chornock}, {Kirshner}, {Leibundgut}, {Dickinson}, {Koekemoer},
  {Grogin}, \& {Giavalisco}}]{2004ApJ...600L.163R}
{Riess}, A.~G., et al. 2004, \apjl, 600, L163

\bibitem[{{Riess} {et~al.}(2007){Riess}, {Strolger}, {Casertano}, {Ferguson},
  {Mobasher}, {Gold}, {Challis}, {Filippenko}, {Jha}, {Li}, {Tonry}, {Foley},
  {Kirshner}, {Dickinson}, {MacDonald}, {Eisenstein}, {Livio}, {Younger}, {Xu},
  {Dahl{\'e}n}, \& {Stern}}]{2007ApJ...659...98R}
{Riess}, A.~G., et al. 2007, \apj, 659, 98

\bibitem[{{Sarkar} {et~al.}(2008){Sarkar}, {Amblard}, {Cooray}, \&
  {Holz}}]{2008ApJ...684L..13S}
{Sarkar}, D., {Amblard}, A., {Cooray}, A., \& {Holz}, D.~E. 2008, \apjl, 684,
  L13

\bibitem[{{Sauer} {et~al.}(2008){Sauer}, {Mazzali}, {Blondin}, {Filippenko},
  {Benetti}, {Stehle}, {Challis}, {Kirshner}, \& {Li}}]{2008arXiv0803.0871S}
{Sauer}, D.~N., et al.
  2008, ArXiv e-prints, 803

\bibitem[{{Scannapieco} \& {Bildsten}(2005)}]{2005ApJ...629L..85S}
{Scannapieco}, E., \& {Bildsten}, L. 2005, \apjl, 629, L85

\bibitem[{{Sullivan} {et~al.}(2006){Sullivan}, {Le Borgne}, {Pritchet},
  {Hodsman}, {Neill}, {Howell}, {Carlberg}, {Astier}, {Aubourg}, {Balam},
  {Basa}, {Conley}, {Fabbro}, {Fouchez}, {Guy}, {Hook}, {Pain},
  {Palanque-Delabrouille}, {Perrett}, {Regnault}, {Rich}, {Taillet}, {Baumont},
  {Bronder}, {Ellis}, {Filiol}, {Lusset}, {Perlmutter}, {Ripoche}, \&
  {Tao}}]{2006ApJ...648..868S}
{Sullivan}, M., et al. 2006, \apj, 648, 868

\end{thebibliography}
\end{document}